\documentclass{ws-mpla}

\begin{document}

\markboth{Thiago Prud\^encio \& Diego Julio Cirilo-Lombardo}
{Coherent states, superpositions and uncertainty relations on a M\"obius strip}

%%%%%%%%%%%%%%%%%%%%% Publisher's Area please ignore %%%%%%%%%%%%%%
\catchline{}{}{}{}{}
%%%%%%%%%%%%%%%%%%%%%%%%%%%%%%%%%%%%%%%%%%%%%%%%%%%%%%%%%%%%%%%%%%%

\title{COHERENT STATES, SUPERPOSITIONS OF COHERENT STATES AND UNCERTAINTY RELATIONS ON A M\"OBIUS STRIP}

\author{\footnotesize THIAGO PRUD\^ENCIO}

\address{Institute of Physics, University of Brasilia - UnB, CP: 04455, 70919-970, Brasilia - DF, Brazil.
thprudencio@gmail.com}

\author{DIEGO JULIO CIRILO-LOMBARDO}

\address{International Institute of Physics - UFRN, Natal - RN, Brazil.\\
Bogoliubov Laboratory of Theoretical Physics, Joint Institute for Nuclear Research 141980,
Dubna-Moscow, Russia.}

\maketitle

\pub{Received (Day Month Year)}{Revised (Day Month Year)}

\begin{abstract}
Since symmetry properties of coherent states (CS) on M\"obius strip (MS) and fermions are closely 
related, CS on MS are naturally associated to the topological properties of fermionic fields. Here we consider CS 
and superpositions of coherent states
(SCS) on MS. We extend a recent propose of CS on MS (Cirilo-Lombardo, 2012 [25]), including the analysis of periodic behaviors of CS and SCS on MS
and the uncertainty relations associated to angular momentum and the phase angle. 
The advantage of CS and SCS on MS with respect to the standard ones and potential applications in continuous variable
quantum computation are also addressed.
\end{abstract}

\ccode{PACS Nos.: 03.70.+k,04.50.-h,42.50.Gy}

The uncertainty relation between amplitude and phase is the minimum 
allowed by the Heisenberg's uncertainty principle in the case of coherent states (CS) in quantum optics \cite{walls}, for this 
reason superpostions of coherent states (SCS) are the best analogues of what are called cat states \cite{schrodinger}. Such 
states are experimentally achievable, for instance, in electrodynamic cavities \cite{brune,raimond} and systems of trapped ions 
\cite{monroe}. The main experimental difficult for 
the creation and observation of such superpositions is related to the fast decay of 
the coherences \cite{huyet}. Advances in continuous variable quantum computation (CVQC) \cite{cerf,neegard} have lead to the possibility 
of manipulation of such states as qubit states and realization of
gate operations \cite{tipsmark}. Although SCS can appear in different contexts 
\cite{cirilo,provost,gazeau2,klauder,barut,olmo}, in nontrivial topologies it is not clear if SCS are always 
associated to cat states \cite{peremolov}. Cat states associated to 
topological defects have been proposed recently \cite{zurek}. 

On the other hand, SCS states and the relation to cat states have not been investigated in M\"obius strip (MS) topology. 
Proposes involving physics in MS as applications in persistent currents \cite{yakubo}, 
topological insulators \cite{sun,lee} and graphene \cite{wang} have increased the interest in the study 
of states and dynamics in MS. The study of CS on MS has been proposed recently in \cite{diego}, extending the torus and 
circle cases \cite{kowalski,kastrup}. A natural step further in this direction is consider SCS states and the uncertainty 
relations in MS in order to realize cat states on MS, whose applications in CVQC can motivate CS and topological 
quantum computation in physical systems with MS topologies.

In this paper, we consider CS and SCS on MS, the relation between SCS and
cat states on MS and the uncertainty relations for CS and SCS on MS. We start with a CS on MS, paremetrized from the torus, by means 
of an angular contraint introduced in \cite{diego}, then we consider the
 uncertainty relations proposed in \cite{kowalsky3} extended to MS case. Finally, we realize minimal uncertainties for
 CS and SCS on MS, that can be associated to cat states on MS.

\section{CS and SCS on M\"obius strip}

\begin{figure}[]
\centering
\includegraphics[scale=0.45]{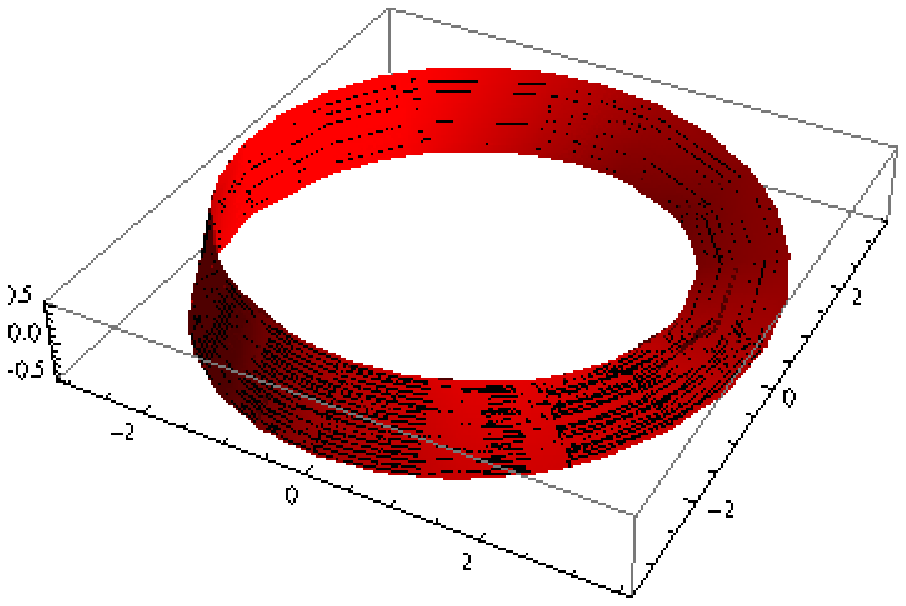}
\includegraphics[scale=0.45]{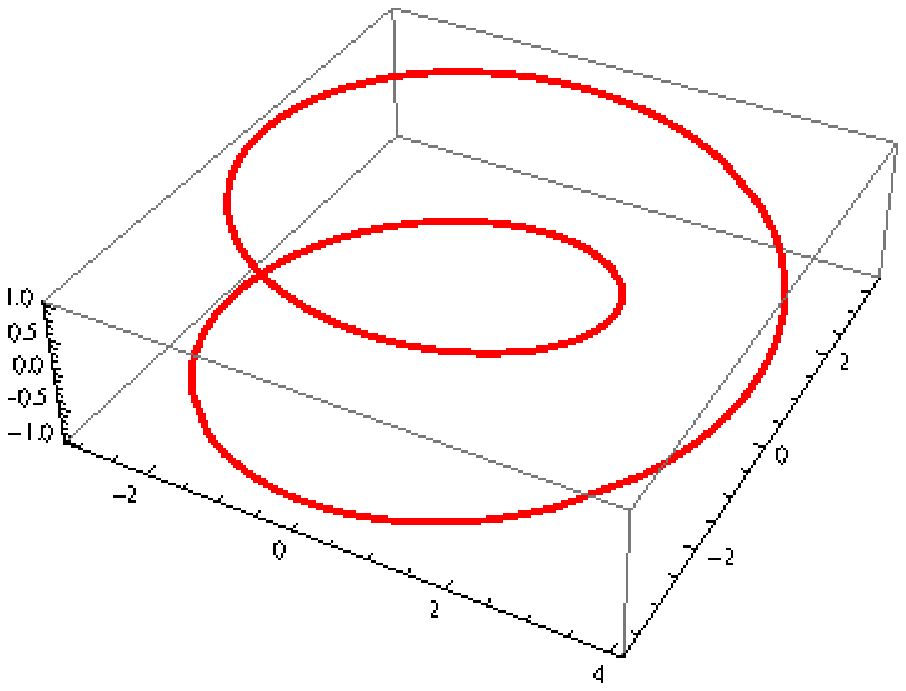}
\caption{(Color online) M\"obius strip and a particular trajectory of a point particle on a M\"obius strip.}
\label{mobius_line}
\end{figure}

Considering a phase space embedded into a torus surface, with an angular constraint between
polar $\theta$ and azimuthal $\varphi$ angles given by $\theta =\frac{\varphi +\pi }{2}$, we have an effectively reduced degree of freedom from the torus to
the MS (figure \ref{mobius_line}), with a trajectory described by the parametrization 
\begin{eqnarray}
X &=& \cos\varphi + r \cos\left(\varphi/2\right) \cos\varphi \nonumber\\
Y &=& \sin\varphi + r\cos\left(\varphi/2\right) \sin\varphi \nonumber\\
Z &=& l+r\sin\left(\varphi/2\right),  
\end{eqnarray}
and the variation of $r < R=1$ leading to the MS. 
The CS for a quantum particle on the
MS is then performed following the eigenvalue equation \cite{diego} 
$e^{i(\widehat{\varphi}+i\widehat{J})}\left\vert \xi\right\rangle =\xi\left\vert \xi\right\rangle$,
where $\xi =e^{-\left( l+r\sin\left( \varphi /2\right) \right) +i\varphi }\left(
1+r\cos\left( \varphi /2\right) \right)$. In terms of $\varphi $ and $l'$ in the MS parametrization, 
the CS can be described by means of 
$|l',\varphi \rangle$, that satisfies $\hat{J}|l',\varphi \rangle =j|l',\varphi \rangle$. In this form, 
the CS on the MS can be written as
\begin{eqnarray}
|\xi\rangle=|l',\varphi\rangle = \sum_{j=-\infty}^{\infty} e^{(l'-i\varphi)j}e^{-j^{2}/2}|j\rangle, 
\end{eqnarray}
where $l'=l+r\sin\left(\frac{\varphi}{2}\right) -\ln\left(1+r\cos\frac{\varphi}{2}\right)$. The normalization
is given in terms of Jacobi theta functions
$\langle \xi|\xi\rangle = \sum_{j,k} e^{(j^{2}+k^{2})/2}e^{(l'-i\varphi)j}e^{(l'+i\varphi)k}\langle k|j\rangle
= \sum_{j}e^{-j^{2}}e^{2l'j} = \theta_{3}(l'|i\pi)$, and the the projections relations 
for two different CS on a MS can be written as
\begin{eqnarray}
\langle \xi|\tilde{\xi}\rangle=\sum_{j=-\infty}^{\infty}e^{(l^{\prime }+%
\tilde{l}^{\prime })j}e^{-i(\varphi-\tilde{\varphi})}e^{-j^{2}}. \label{projection}
\end{eqnarray}
Then, on a period in the MS, we have the behaviour characteristic of the topology associated to 
the MS. Taking $|\xi\rangle=|\xi^{\varphi=0}\rangle$ and $|\tilde{\xi}\rangle=|\xi^{\varphi=4\pi}\rangle$, the projections are given 
by
\begin{eqnarray}
\langle \xi^{\varphi=0}|\xi^{\varphi=4\pi}\rangle %&=&\sum_{j=-\infty}^{\infty}e^{2lj + \ln(1+r)j}e^{-j^{2}}\nonumber\\
= \sqrt{\pi}\exp{(l +\frac{1}{2}\ln(1+r))^{2}}\theta_{3}\left(-\frac{(2l + \ln(1+r))\pi}{2})|e^{-\pi^{2}}\right),
\end{eqnarray}
that can be 
simplified, using (\ref{projection}), in the following form
\begin{eqnarray}
\langle \xi^{\varphi=0}|\xi^{\varphi=4\pi}\rangle = e^{4\pi i} \sum_{j=-\infty}^{\infty}e^{(l^{\prime }+%
\tilde{l}^{\prime })j}e^{-j^{2}}. 
\end{eqnarray}
Since the angular difference is a unity, i.e., $\exp{4\pi i}=1$, we have that the state with a difference in the phase $\varphi$ corresponding to the period $4\pi$
of a MS will be the same $|\xi^{\varphi=0}\rangle$, as expected from the MS topology. Consequently, under a period of
$4\pi$ the state turns to be the same state.  

We can verify that this is in fact a property of SCS too. Considering a SCS given in terms of CS of opposite
angles $\varphi$ and $-\varphi $, including an additional phase term phase $e^{-i\phi}$, we have 
\begin{eqnarray}
|\phi_{C}\rangle&=& |l',\varphi\rangle + e^{-i\phi}|l',-\varphi\rangle. \label{abdc23}
\end{eqnarray}
%In terms of a fiducial vector $|0,0\rangle$ in $r=0$, we can also write
%$|\phi_{C}\rangle = e^{\lbrace [l + r\sin(\varphi/2) -\ln(1+r\cos(\varphi/2))-i\varphi]j \rbrace}|0,0\rangle
%+ e^{-i\phi}e^{\lbrace[l - r\sin(\varphi/2) -\ln(1+r\cos(\varphi/2)) + i\varphi]j\rbrace}|0,0\rangle$. 
As the CS,  the SCS 
(\ref{abdc23}) is periodic in the MS, taking $|\phi_{C}^{\varphi=0}\rangle$ and $|\phi_{C}^{\varphi=4\pi}\rangle$, 
a period of $4\pi$, we can 
write $|\phi_{C}^{\varphi=0}\rangle = |\phi_{C}^{\varphi=4\pi}\rangle$.

Now, if we relate the phase term $e^{-i\phi}$ directly to the MS angular variable $\varphi$, by means
of $\phi=\varphi$, the states will still be periodic on MS, under a period of $4\pi$.

We can also consider a SCS given in terms of opposite CS states $|\xi\rangle$ and $|-\xi\rangle$
\begin{eqnarray}
|\psi_{C}\rangle= |\xi\rangle + e^{i\phi}|-\xi\rangle.
\end{eqnarray}
On a period in the MS, the SCS turns to be the same state. As the states
 $|\pm\xi^{\varphi=0}\rangle=|\pm\xi^{\varphi=4\pi}\rangle$, the SCS 
$|\psi_{C}\rangle$ is also periodic under $4\pi$ period on MS. We can also verify 
that the state $|\psi_{C}\rangle$ is also periodic in the case $\phi=\varphi$. 

The action of the operator $e^{i\hat{\varphi}+i\hat{J}}$ on $|\psi_{C}\rangle$, leads to a SCS also MS periodic, $4\pi$, 
$|\psi_{C}^{-}\rangle=|\xi\rangle - e^{i\phi}|-\xi\rangle$, by means of $e^{i\hat{\varphi}+i\hat{J}}|\psi_{C}\rangle=\xi|\psi_{C}^{-}\rangle$ .

\begin{figure}[]
\centering
\includegraphics[scale=0.45]{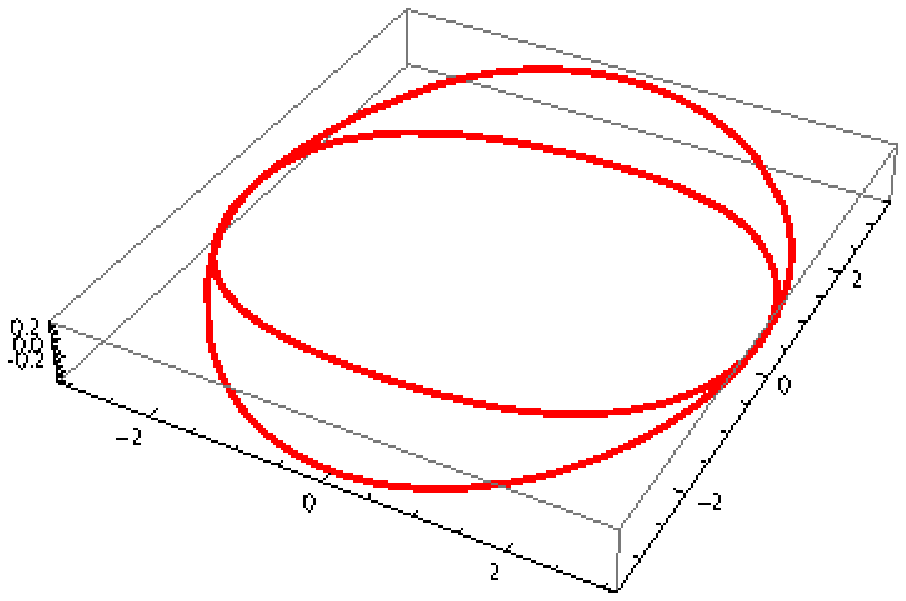}
\includegraphics[scale=0.45]{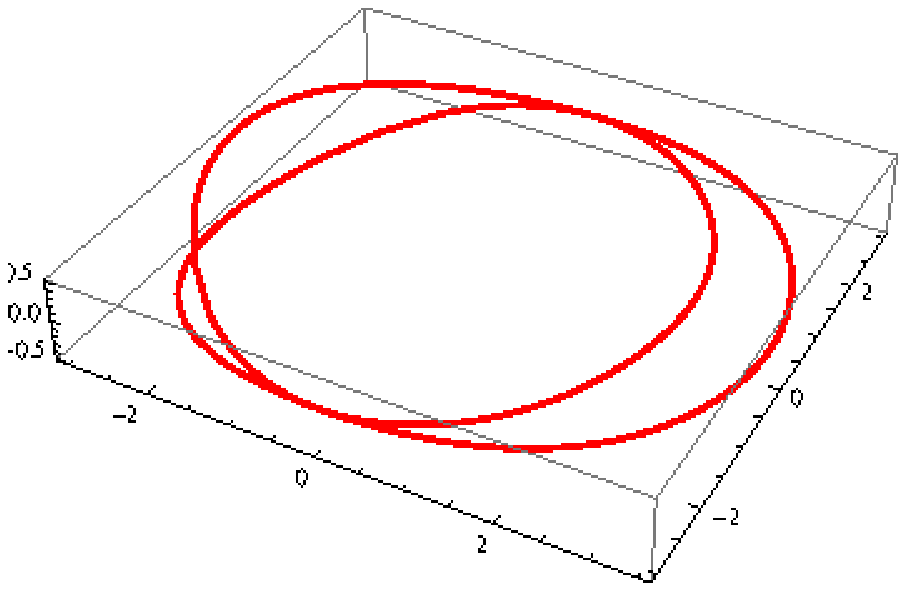}
\caption{(Color online) Trajectories varying $r$ along the MS, $r=\sin^{2}(\varphi)/2$ and $r=\cos^{2}(\varphi)/2$.}
\label{mobius22}
\end{figure}

If $r$ changes along the MS, the periodicity is still MS dependent. But the particle can realize an harmonic motion. For instance, 
$r=\sin^{2}(\varphi)/2$, the SCS carries a circle periodicity of $2\pi$ on the MS, with
 $|\psi_{C}^{\varphi=0}\rangle=|\psi_{C}^{\varphi=2\pi}\rangle=|\psi_{C}^{\varphi=4\pi}\rangle$. On the other hand, 
for $r=\cos^{2}(\varphi)/2$ the SCS carries a periodicity of $4\pi$ on the MS (figure \ref{mobius22}). In the case of $r=0$, that corresponds to a cilinder parametrization 
$X = \cos\varphi$, $Y = \sin\varphi$, $Z = l$, the SCS turn to the
same state after a period of $2\pi$. Then, variations of $r$ in the MS generally do not alter the periodic properties of SCS, leading to circle or MS periodicities. This is 
an important point for the study the stability of SCS in the MS.

\section{Uncertainty Relations}

CS provides a close connection between quantum and classical physics, 
with a suitable set of requirements, as continuity, resolution of unity, 
temporal stability and action identity. In particular, these properties are reflected in SCS, that naturally 
will lead to cat states, the best analogues 
of classical behaviour. 

SCS with minimal uncertainty relations can be associated to cat states, that are the best analogues 
of classical superpositions. In order to achieve them on MS, we have to
consider the minimization of uncertainty relation in the context of MS.
 
The uncertainty measurements associated to angular momentum $\hat{J}$ and the phase angle $\hat{\varphi}$, written 
in the form proposed by \cite{kowalsky3}, for the MS case are given by
\begin{eqnarray}
\Delta^{2}_{\xi}(\hat{J})&=& \frac{1}{4}\left|
\ln\left(\langle e^{-2\hat{J}}\rangle_{\xi}\langle e^{2\hat{J}}\rangle_{\xi}\right)\right|, \\
\Delta^{2}_{\xi}(\hat{\varphi})&=& \frac{1}{4}\left|\ln\left(\frac{1}{|\langle e^{2i\hat{\varphi}} \rangle_{\xi}|^{2}}\right)\right|.
\end{eqnarray}
where the expectation value of $e^{i\hat{\varphi}}$, associated to the CS $|\xi\rangle$ on MS, can 
be $\langle \xi |e^{i\hat{\varphi}}|\xi\rangle=\sum_{j} e^{l'+i\varphi}e^{-1/2}\sum_{j}e^{-j(j+1)}e^{2j'l'}$, that can 
be simplified in terms of 
%\begin{eqnarray}
%\langle \xi |U|\xi\rangle &=& \sum_{j,k}e^{(l'+i\varphi)j}e^{(l'+i\varphi)k}e^{-(j^{2}+k^{2})/2}\langle k|j+1\rangle \nonumber \\
%&=& \sum_{j} e^{(l'-i\varphi)j}e^{(l'+i\varphi)j}e^{-(2j^{2}+1+2j)/2} \nonumber \\
%&=& \sum_{j} e^{l'+i\varphi}e^{-1/2}\sum_{j}e^{-j(j+1)}e^{2j'l'}
%\end{eqnarray}
Jacobi theta functions $\langle \xi|U|\xi\rangle = e^{(l'+i\varphi)}e^{-1/2}\theta_{3}(l'-\frac{1}{2}|i\pi)$. 
The same procedure applied to $e^{2i\hat{\varphi}}$, leads to 
$\langle \xi|e^{2i\hat{\varphi}}|\xi\rangle = e^{2(l'+i\varphi)}e^{-2}\sum_{j}e^{2l'j}e^{-2(j+1)}e^{-j^{2}}$, that, in terms of Jacobi theta 
functions is given by 
$\langle \xi|e^{2i\hat{\varphi}}|\xi\rangle=e^{2(l'+i\varphi)}e^{-2}\theta_{3}(l'-1|i\pi)$.
%\begin{eqnarray}
%\langle \xi|U^{2}|\xi\rangle &=& \sum_{j,k} e^{(l'-i\varphi)j}e^{-j^{2}/2}e^{-k^{2}/2}e^{(l'+i\varphi)k}\langle k|j+2\rangle \nonumber \\
%&=& \sum_{j}e^{(l'-i\varphi)j}e^{(l'+i\varphi)j}e^{(l'+i\varphi)2}e^{-j^{2}/2}e^{-(j^{2}+4j+4)/2} \nonumber \\
%&=& e^{2(l'+i\varphi)}\sum_{j}e^{2l'j}e^{-j^{2}}e^{-2(j+1)} \nonumber \\
%&=& e^{2(l'+i\varphi)}e^{-2}\sum_{j}e^{2l'j}e^{-2(j+1)}e^{-j^{2}} \label{u2}
%\end{eqnarray}
%If we write the above expression in terms of Jacobi theta function, we have
Dividing both results by the CS normalization, we arrive at 
\begin{eqnarray}
\langle e^{2i\hat{\varphi}} \rangle_{\xi_{MS}} = e^{2(l' + i\varphi)}e^{-2}\frac{\theta_{3}(l'-1|i\pi)}{\theta_{3}(l'|i\pi)}, \label{gen1} \\
\langle e^{i\hat{\varphi}} \rangle_{\xi_{MS}} =
 e^{(l'+i\varphi)}e^{-1/2}\frac{\theta_{3}(l'-\frac{1}{2}|i\pi)}{\theta_{3}(l'|i\pi)},
\end{eqnarray}
For the operators $e^{2\hat{J}}$ and $e^{-2\hat{J}}$ 
the expectation value associated to the CS on MS, that will lead to 
$\langle e^{-2\lambda \hat{J}}\rangle_{\xi_{MS}}= \langle \xi_{MS}|e^{-2\lambda \hat{J}}| \xi_{MS} \rangle/\langle \xi_{MS}| \xi_{MS} \rangle$,
 can be calculated as in $e^{i\hat{\varphi}}$ case, leading to
\begin{eqnarray}
\langle e^{-2\lambda \hat{J}}\rangle_{\xi_{MS}} &=& e^{\lambda^{2}-2l\lambda}e^{-2\lambda[r\sin\varphi/2]}\nonumber \\
&\times&(1+r\cos(\frac{\varphi}{2}))^{2\lambda}\frac{\theta_{3}(l'-\lambda|i\pi)}{\theta_{3}(l'|i\pi)},
\end{eqnarray}
where $\lambda=\pm 1$.
\begin{figure}[]
\centering
\includegraphics[scale=0.4]{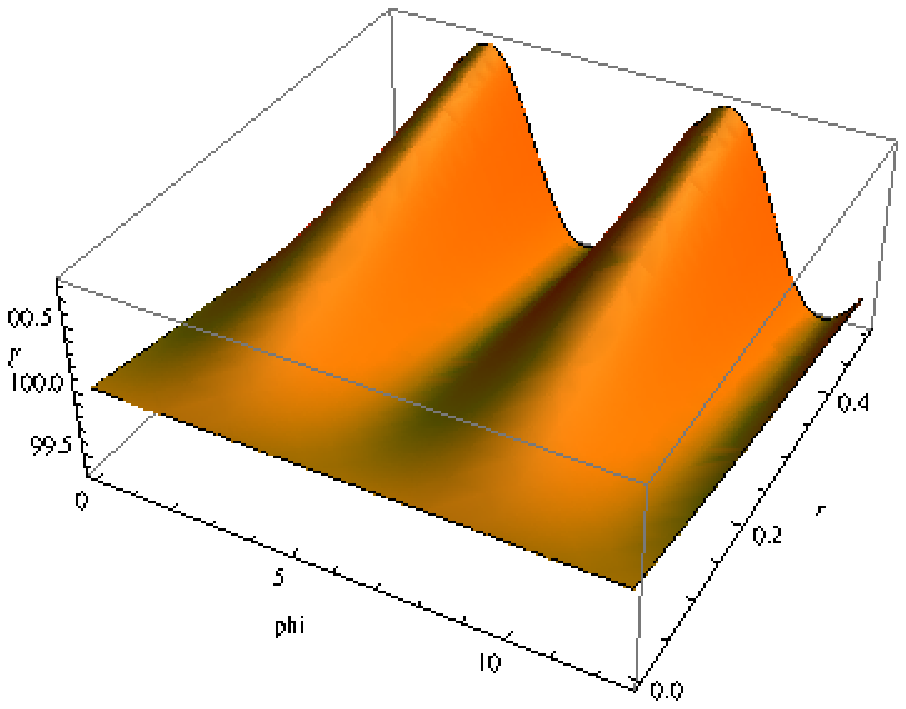}
\includegraphics[scale=0.4]{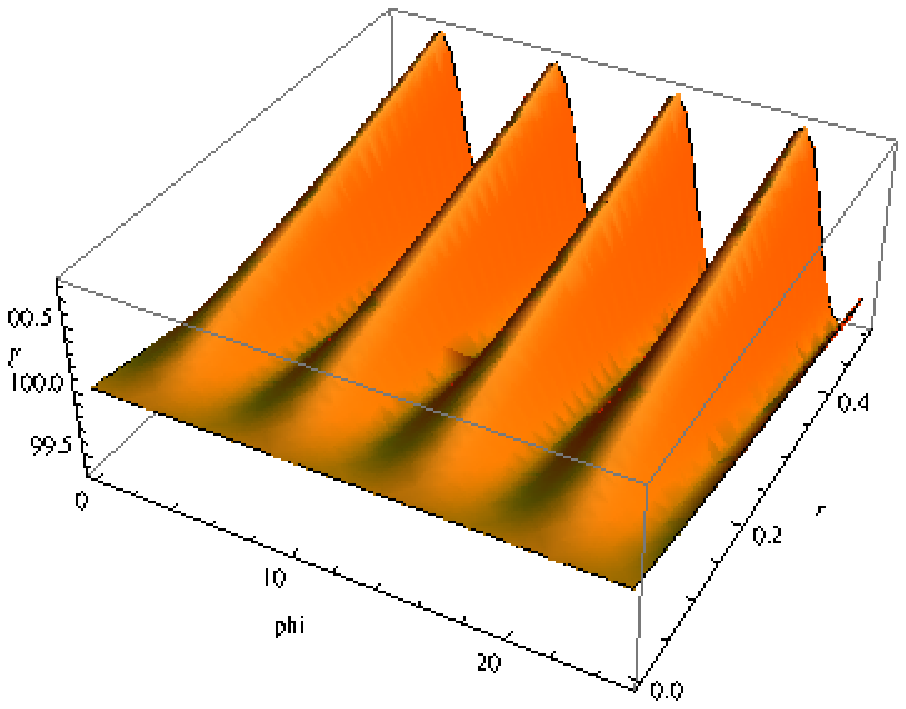}
\caption{(Color online)Uncertainty measurements associated to angular momentum and the phase angle for 
(a) $l'$ in a period of $2\pi$, (b) $l'$ in a period of $4\pi$.}
\label{lprima}
\end{figure}
For sufficiently large $l'$, we have
\begin{eqnarray}
\Delta^{2}_{\xi_{MS}}(\hat{J}) + \Delta^{2}_{\xi_{MS}}(\hat{\varphi}) = l', \label{ur}
\end{eqnarray}
that reduces to the circle case for $r=0$. In particular, this uncertainty relation has a set of minimum values depending 
on $l'$ along a MS period $4\pi$ in the angular variable (figure \ref{lprima}). 

Taking into account the uncertainty relation in a more usual form \cite{sakurai} 
$\langle (\Delta \hat{A})^{2}\rangle\langle (\Delta \hat{B})^{2}\rangle \geq \frac{1}{4}\left|\langle[\hat{A},\hat{B}]\rangle\right|^{2}$, for the case 
of $\hat{J}$ and $e^{i\hat{\varphi}}$, that satisfy the commutation relation
 $[\hat{J},e^{i\hat{\varphi}}]=e^{i\hat{\varphi}}$ on MS \cite{diego}, we can also write
\begin{eqnarray}
\langle (\Delta \hat{J})^{2}\rangle\langle (\Delta e^{i\hat{\varphi}})^{2}\rangle \geq \frac{1}{4}\left|\langle e^{i\hat{\varphi}}\rangle\right|^{2}
\end{eqnarray}
that using $\langle e^{i\hat{\varphi}} \rangle_{\xi_{MS}}$ we find that the minimum uncertainty relations on MS 
satisfy
\begin{eqnarray}
\langle (\Delta \hat{J})^{2}\rangle\langle (\Delta e^{i\hat{\varphi}})^{2}\rangle 
= \frac{1}{4}\left|\left[e^{(l'+i\varphi)}e^{-1/2}\frac{\theta_{3}(l'-\frac{1}{2}|i\pi)}{\theta_{3}(l'|i\pi)}\right]\right|^{2}. \nonumber \\
\end{eqnarray}
For sufficiently large $l'$, we have $\left|\langle e^{i\hat{\varphi}}\rangle_{\xi_{MS}}\right|^{2} = e^{2l'},$ 
and the corresponding uncertainty relation is given by $e^{2l'}/4$; 
for $l'=0$, $|\langle e^{i\hat{\varphi}} \rangle_{\xi_{MS}}|^{2}/4 = e/4$. For a small phase angle we can also write
\begin{eqnarray}
\langle (\Delta \hat{J})^{2}\rangle\langle (\Delta \hat{\varphi})^{2}\rangle \geq \frac{1}{4}.
\end{eqnarray}

As a consequence, SCS associated to minimum uncertainty relation states are directly related to cat states on MS, 
these states are important for the study of classical analogues on MS and the possibility of CS quantum computation on MS. Since the CS and 
SCS constructions on MS can be also reduced to circle topology cases, by $r=0$ or some variations of $r$ along the MS, this CS states are 
quite general. 

In contrast to CS for the fermions where 
 the cylinder topology is used, forcing to introduce spin-1/2 
by hand, the CS on MS is a natural framework phase space where the $4\pi$ symmetry invariance emerges naturally,
the symmetry properties of MS and fermions are then closely related, both
with characteristic double covering that makes the symmetry invariance of $4\pi$ instead
of $2\pi$ for the bosonic case. Due to
the double covering, CS on MS can also include the bosonic case, as we have showed in $r=0$ 
or an adequate choice of $r$ variation with $\varphi$ angular variable.   
The mininum uncertainty relations for CS and SCS on MS also leads to a natural phase space where
 cat states in a nontrivial topology can be formulated.

\section{Conclusions}

We have considered CS and SCS on MS, the uncertainty measurements associated to angular momentum
 and the phase angle for the MS and the connection 
between SCS and cat states was considered. As a result, we extended previous proposes, including here 
the case of the uncertainty relation in the MS case. In particular, the uncertainty measurements can be associated to MS parameter 
$l'$.

Since the symmetry properties of MS and fermions are closely related, due to 
the double covering property shared by MS and fermions, symmetry invariance of 
$4\pi$, CS on MS naturally describes fermionic fields. Due to
the double covering, CS on MS can also play the role of projector to the bosonic case \cite{kempf}, making a closer connection between
classical and quantum formulations \cite{gazeau} in a Dirac-like quantization \cite{diego}. When the 
particle is confined to a M\"obius topology its periodicity is similar to a fermionic
behavior of a spin $1/2$ particle and the similarities can be used to investigate the couterparts
associated to cat states in these systems, but it is important to emphasize that the $4\pi$ invariance due 
to the MS topology is an invariance in the physical space,
while the $4\pi$ invariance associated with fermions is an invariance in the internal spin space.
The behavior of CS and SCS on MS are associated to a continous transformation, in contrast to the discrete behaviour of the spin
system, but analogue to this one under spin rotations. 

CS can be very sensitive to the geometry they are constructed, the CS on MS has the advantage of recover cylinder topologies 
as particular cases when $r=0$ or an adequate choice of $r$ variation with $\varphi$ angular variable. As MS is 
a more natural phase space for fermions symmetries of spin rotation, they can also be useful for investigating cat states analogues to 
fermionic fields.
 
SCS on MS are strong canditates to cat states in nontrivial topologies. This is 
reinforced by the minimum uncertainty relations for CS on MS. One consequence is the possibility of using
 CS and SCS on MS to realize CVQC in nontrivial topologies.

For the SCS in a period on MS, we showed the expected behaviour of periodicity in phase $\varphi$ with a period 
of $\varphi =4\pi$ in the MS. In the case $r=0$, corresponding to the case in the torus, 
the periodicity is, as expected, of $\varphi =2\pi$. By varying $r$ 
along the MS, the particle can describe an oscillatory motion along the strip
such that there is a mixing of the cases in the torus and MS cases. As the CS comes from a constraint on the torus parametrization, 
the CS on MS can also be connected to the CS on torus by means of the breaking of constraint between the angular variables. 
The topological stability of SCS, as showed for particular $r$ variations, can also be used 
for CS quantum computation, topological physics on MS and formulations on Riemannian superspaces \cite{alfonso}. 

\section{Acknowledgements}
DJCL thanks CNPQ (Brazil) for financial support. TP thanks CAPES (Brazil) for partial financial support.
The authors are also grateful for the reviewers comments and suggestions that have made this work more presentable.


\begin{thebibliography}{99}
\bibitem{walls} D. F. Walls, G. J. Milburn, \textit{Quantum Optics} (Springer-Verlag, Berlin, 2008).
\bibitem{schrodinger} E. Schrodinger, Naturwissenschaften, \textbf{23} (1935) 807.
\bibitem{brune} M. Brune, E. Hagley, J. Dreyer, X. Matre, A. Maali, C.
Wunderlich, J.M. Raimond, and S. Haroche, Phys. Rev. Lett. \textbf{77}, (1996) 4887.
\bibitem{raimond} J. M. Raimond, M. Brune, S. Haroche, Rev. Mod. Phys., 
\textbf{73} (2001) 565.
\bibitem{monroe} C. Monroe, D.M. Meekhof, B.E. King, D.J. Wineland,
Science \textbf{272} (1996) 1131.
\bibitem{huyet} G. Huyet, S. Franke-Arnold, S. M. Barnett, Phys. Rev. A, \textbf{63} (2001) 043812.
\bibitem{cerf} N.
Cerf, G. Leuchs, and E. Polzik, \textit{Quantum Information with
Continuous Variables of Atoms and Light} (Imperial College Press,
London, 2007).
\bibitem{neegard} J. S. Neegard-Nilsen, et al., Phys. Rev. Lett. {\bf 105} (2010) 053602.
\bibitem{tipsmark} A. Tipsmark, et al., Phys. Rev. A, {\bf 84} (2011) 050301.
\bibitem{cirilo} D. J. Cirilo-Lombardo, Phys. Part. Nucl. Lett. {\bf 6} (2009) 359.
\bibitem{provost} J. P. Provost, G. Vallee, Commun. Math. Phys. {\bf 76} (1980) 289.
\bibitem{gazeau2} J.P. Gazeau, Coherent States in Quantum Physics (2009)
\bibitem{klauder} Klauder J R and Skagerstam B S, Coherent States: Applications in Physics and Mathemetical Physics, (1985).
\bibitem{barut} A. O. Barut, L. Girardello, Commun. Math. Phys. \textbf{21} (1971) 41.
\bibitem{olmo} J. A. Gonzalez, M. A. Olmo, J. Phys. A: Math. Gen. \textbf{31} (1998) 8841.
\bibitem{peremolov} A. Perelomov, Generalized Coherent States and their Applications (Springer, Berlin, 1986).
\bibitem{zurek} J. Dziarmaga, W. H. Zurek, M. Zwolak, Nature Physics \textbf{8} (2012) 49.
\bibitem{yakubo} K. Yakubo, Y. Avishai,  D. Cohen, Phys. Rev. B \textbf{67} (2003) 125319. 
\bibitem{kastrup} H. A. Kastrup, Phys. Rev. A \textbf{73} (2006) 052104. 
\bibitem{kowalski} K. Kowalski, J. Rembielinski, Phys. Rev. A, \textbf{75} (2007) 052102.
\bibitem{bayfield} J. E. Bayfield and L. A. Pinnadmvage, Phys. Rev. Lett. \textbf{54} (1985) 313.
\bibitem{sun} Z. L. Guo, Z. R. Gong, H. Dong, C. P. Sun, Phys. Rev. B, \textbf{80} (2009) 195310.
\bibitem{lee} L-T. Huang, D-H. Lee, Phys. Rev. B \textbf{84} (2011) 193106.
\bibitem{wang} X. Wang, X. Zheng, M. Ni, L. Zou, and Z. Zenga, Applied Physics Letters \textbf{97} (2010) 123103.
\bibitem{diego} D. J. Cirilo-Lombardo, J Phys. A: Math. Theor. {\bf 45} (2012) 244026.
\bibitem{klauder2} J. R. Klauder, Phys. Rev. A, \textbf{34} (1986) 4486.
\bibitem{kowalsky} K. Kowalski et al., J.Phys.A, \textbf{29} (1996) 4149.
\bibitem{kowalsky3} K. Kowalski, J. Rembielinski, J. Phys. A: Math. Gen. \textbf{35} (2002) 1405.
\bibitem{sakurai} J. J. Sakurai, \textit{Modern Quantum Mechanics} (Addison-Wesley Publishing Company, Massachusetts, 1994). 
\bibitem{kempf} A. Kempf, J. R. Klauder, J.Phys. A: Math. Gen. \textbf{34} (2001) 1019.
\bibitem{gazeau} J. P. Gazeau, J. R. Klauder, J. Phys. A: Math. Gen. \textbf{32} (1999) 123.
\bibitem{alfonso} D. J. Cirilo-Lombardo, V. I. Afonso, Phys. Lett. A, \textbf{376} (2012) 3599.
\end{thebibliography}
\end{document}